\documentclass[aps, prl, reprint, superscriptaddress]{revtex4-1}

\usepackage{amsmath}
\usepackage{amssymb}
\usepackage{wasysym}
\usepackage{graphicx}
\usepackage{hyperref}
\usepackage{color}
\usepackage{physics}
\usepackage{siunitx}
\usepackage{xcolor}
\usepackage{changes}
\usepackage{multirow}

\allowdisplaybreaks

\begin{document}

\title{Tune-Out and Magic Wavelengths for Ground-State $^{23}$Na$^{40}$K Molecules}

\author{Roman~Bause}
\email{roman.bause@mpq.mpg.de}
\affiliation{Max-Planck-Institut f\"{u}r Quantenoptik, 85748 Garching, Germany}
\affiliation{Munich Center for Quantum Science and Technology, 80799 M\"{u}nchen, Germany}
\author{Ming~Li}
\affiliation{Department of Physics, Temple University, Philadelphia, Pennsylvania 19122, USA}
\author{Andreas~Schindewolf}
\author{Xing-Yan~Chen}
\author{Marcel~Duda}
\affiliation{Max-Planck-Institut f\"{u}r Quantenoptik, 85748 Garching, Germany}
\affiliation{Munich Center for Quantum Science and Technology, 80799 M\"{u}nchen, Germany}
\author{Svetlana~Kotochigova}
\affiliation{Department of Physics, Temple University, Philadelphia, Pennsylvania 19122, USA}
\author{Immanuel~Bloch}
\affiliation{Max-Planck-Institut f\"{u}r Quantenoptik, 85748 Garching, Germany}
\affiliation{Munich Center for Quantum Science and Technology, 80799 M\"{u}nchen, Germany}
\affiliation{Fakult\"{a}t f\"{u}r Physik, Ludwig-Maximilians-Universit\"{a}t, 80799 M\"{u}nchen, Germany}
\author{Xin-Yu~Luo}
\affiliation{Max-Planck-Institut f\"{u}r Quantenoptik, 85748 Garching, Germany}
\affiliation{Munich Center for Quantum Science and Technology, 80799 M\"{u}nchen, Germany}

\date{\today}

\begin{abstract}
We demonstrate a versatile, state-dependent trapping scheme for the ground and first excited rotational states of $^{23}$Na$^{40}$K molecules. Close to the rotational manifold of a narrow electronic transition, we determine tune-out frequencies where the polarizability of one state vanishes while the other remains finite, and a magic frequency where both states experience equal polarizability. The proximity of these frequencies of only 10\,GHz allows for dynamic switching between different trap configurations in a single experiment, while still maintaining sufficiently low scattering rates.
\end{abstract}

\maketitle
Trapping potentials for ultracold atoms and molecules are based on spatially dependent energy shifts of their internal states produced by magnetic, electric, or optical fields. Generally, these energy shifts are state dependent, which greatly affects the time evolution of superposition states of atoms or molecules. Demand in precision quantum metrology, simulation and computation have motivated the careful design of state-dependent traps that offer better control over quantum states. One limiting case is the magic trapping condition, where the light shift of two internal states is identical~\cite{Katori_1999, Kotochigova_2010, Neyenhuis_2012}. It is a key ingredient for achieving long coherence time in atomic and molecular clocks~\cite{Ye_2008, Kondov_2019, Leung_2020}. Another limiting case is the tune-out condition, where the light shift of one state vanishes while the other remains finite~\cite{LeBlanc_2007, Arora_2011}. Such highly state-dependent potentials can be used in novel cooling schemes for atoms~\cite{Catani_2009}, selective addressing and manipulation of quantum states~\cite{Kotochigova_2006, Wang_2015b, Rubio_Abadal_2019}. Tune-out wavelengths have also been used for precision measurements of atomic structure~\cite{Holmgren_2012, Herold_2012, Petrov_2013, Henson_2015, Kao_2017, Heinz_2019}.

We extend these concepts to rotational states of ultracold polar molecules~\cite{Ni_2008, Takekoshi_2014, Molony_2014, Park_2015a, Guo_2016, McCarron_2018, Rvachov_2017, Yang_2019, Liu_2019, Anderegg_2019}. Such molecules offer unique possibilities for quantum engineering due to their strong long-range dipolar interactions and long single-particle lifetime~\cite{Carr_2009, Baranov_2012, Kwasigroch_2017, Bohn_2017, Ni_2018}. Manipulating their rotational degrees of freedom is particularly important for experimental control of dipolar interactions. Though significant advances in controlling the internal states of molecules have been made~\cite{Ospelkaus_2010, Yan_2013,  Gregory_2016, Will_2016, Prehn_2016, Guo_2018a, Caldwell_2019}, engineering rotational states in optical dipole traps remains technically challenging. This is due to both the complex level structure of molecules and the strong anisotropic coupling between the rotation of molecules and optical trapping fields. In far-detuned optical dipole traps, rotational magic conditions only exist at special light polarizations or intensities~\cite{Neyenhuis_2012, Gregory_2017, Seesselberg_2018b, Blackmore_2018, Rosenband_2018}. This results in a high sensitivity to polarization or intensity fluctuations, which limits rotational coherence times. Additionally, tune-out conditions, which could be powerful tools for evaporative cooling in optical lattices, have never been demonstrated for polar molecules. In this work, we demonstrate a versatile, rotational-state dependent trapping scheme by using laser light near-resonant with rotational transition lines of a nominally forbidden molecular transition. This allows us to create tune-out and magic conditions for rotational states of molecules by controlling the laser frequency, with first- and second-order insensitivity to the polarization angle and intensity of light.

\begin{figure}
\centering
\includegraphics{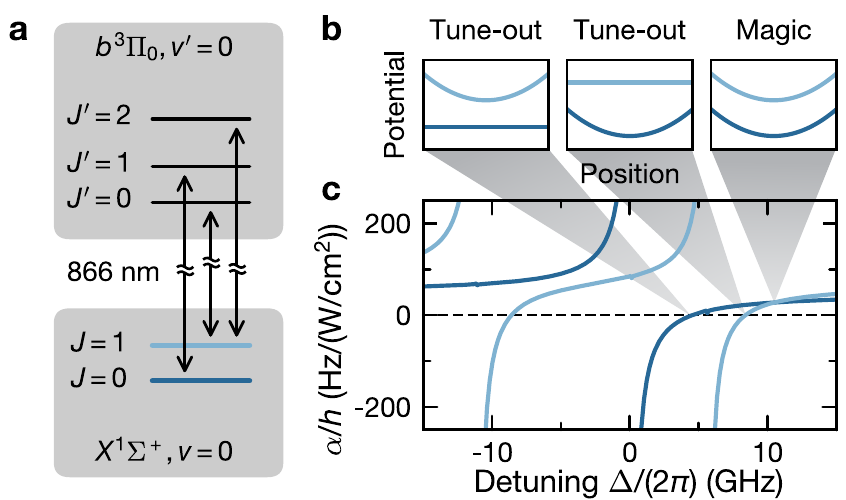}
\caption{Overview of the rotational-state dependent trapping scheme near the $X \leftrightarrow b$ transition. (a) Level diagram of the NaK molecule containing the $X \leftrightarrow b$ transition and the two nearest transitions from $|1 \rangle$. (b) Schematic depiction of the potential experienced by $| 0 \rangle$ (dark blue) and $| 1 \rangle$ (bright blue) molecules in a dipole trap at the tune-out detuning for $|0\rangle$ (left panel), a tune-out detuning for $|1\rangle$ (center panel) and the magic detuning (right panel). (c) Frequency-dependent polarizability for $|0\rangle$ (dark blue) and $|1\rangle$ (bright blue), assuming light polarization parallel to the quantization axis. Each pole corresponds to one of the transitions shown in (a).}
\label{fig:fig1}
\end{figure}

In our experiments, we use $^{23}$Na$^{40}$K molecules in their rovibrational ground state $| X^1 \Sigma^+, v=0, J=0 \rangle $ as well as their first rotationally excited state, $|J=1, m_J=0 \rangle$. In the following, we will refer to these states as $|0 \rangle$ and $|1 \rangle$, respectively. The rotational-state dependent dipole trap is realized with laser light slightly detuned from the $| X^1 \Sigma^+, v=0, J=0 \rangle \leftrightarrow | b^3 \Pi_0, v^\prime=0, J^\prime=1 \rangle$ transition (subsequently called the $X \leftrightarrow b$ transition), which was previously studied in \cite{Kobayashi_2014, Harker_2015}. For detunings from this transition comparable to the rotational constants, dynamic polarizabilities depend strongly on the rotational level of the $X$ state (see Fig.\ \ref{fig:fig1}). Tune-out conditions for both states as well as a magic condition can thereby be achieved within a frequency range of less than 10 GHz. All intermediate ratios of polarizability can be realized between these limiting cases. The $X \leftrightarrow b$ transition is mostly electric-dipole forbidden and therefore exhibits a narrow partial linewidth of $\Gamma = 2\pi\times \SI{301(10)}{\hertz}$, which is a measure of the decay rate from an initial state to a specific final state. This value is much smaller than the spacing between rotational states, which leads to photon scattering rates small enough to realize dipole traps at the tune-out and magic frequencies.

The frequency of the $X \leftrightarrow b$ transition is $\omega_0 = 2 \pi \times \SI{346.12358(7)}{\tera \hertz}$, corresponding to a wavelength of $\lambda = \SI{866.1428(3)}{\nano \meter}$. The polarizabilities $\alpha_{0}(\Delta)$ and $\alpha_{1}(\Delta)$ of a molecule in $|0\rangle$ or $|1\rangle$, respectively, in a light field detuned by $\Delta$ from the $X \leftrightarrow b$ transition can be described by 
\begin{eqnarray}
\label{polarizability_0}
\alpha_0 &=& -\frac{3 \pi c^2}{2 \omega_0^3} \frac{\Gamma}{\Delta} + \alpha_{\mathrm{iso}}, \\
\label{polarizability_1}
\nonumber \alpha_1 &=& -\frac{3 \pi c^2}{2 \omega_0^3} \left( \frac{\Gamma \cos^2\theta}{\Delta + 2(B+B^\prime)/\hbar} \right. \\  & & \left. + \frac{1}{5}\frac{\Gamma(\cos^2\theta + 3)}{\Delta - 2(2B^\prime - B)/\hbar} \right) + \alpha_{\mathrm{iso}} + \alpha_{\text{ang}}(\theta)
\end{eqnarray}
in the case where $\Delta$ is much larger than the hyperfine structure of the resonance. Here, $\alpha_{\mathrm{iso}}$ and $\alpha_{\text{ang}}(\theta)$ are background terms that describe the polarization-independent and -dependent contributions from the other far-detuned transitions, respectively, $\theta$ denotes the angle between the light polarization and the quantization axis, which is given by the direction of the dc electric field in the experiment. $B$ and $B^\prime$ denote the ground and excited-state rotational constants, respectively. The background polarizability terms can be expressed as~\cite{Kotochigova_2010, Li_2017, Seesselberg_2018b}
\begin{eqnarray}
\label{background_pol_iso}
\alpha_{\mathrm{iso}} &=& \frac{1}{3} \left(\alpha^\parallel_{\mathrm{bg}} + 2\alpha^\perp_{\mathrm{bg}} \right), \\
\label{background_pol_ang}
\alpha_{\text{ang}} &=& \frac{2}{15}\left[ 3 \cos^2 ( \theta) -1 \right] (\alpha^\parallel_{\mathrm{bg}} - \alpha^\perp_{\mathrm{bg}}),
\end{eqnarray}
where $\alpha^\parallel_{\mathrm{bg}}$ and $\alpha^\perp_{\mathrm{bg}}$ are the background parallel and perpendicular polarizabilities, respectively. The photon scattering rate of molecules in $|0\rangle$ near the $X \leftrightarrow b$ transition is given by 
\begin{equation}
\label{scattering-rate}
\gamma_{\text{sc}} = \frac{3 \pi c^2}{2 \hbar \omega_0^3} \frac{\Gamma \Gamma_e}{\Delta^2}I,
\end{equation}
where $I$ is the light intensity and $\Gamma_e$ is the natural linewidth of the excited state. 

Our experimental cycle begins with the preparation of a near-degenerate sample of molecules in the $|0\rangle$ state using STIRAP, as described in~\cite{Seesselberg_2018a}. Depending on the measurement, this preparation is done either in a far-detuned crossed-beam optical dipole trap or a one- or three-dimensional (1D or 3D) optical lattice, see Supplemental Material \cite{supplement}. All our measurements are performed at a magnetic field of \SI{85.4}{G}. The $1/e$ radius of the molecule cloud is $\approx$ 30$\, \mu$m. In order to image the molecules, we perform a reverse STIRAP procedure and employ absorption imaging on molecules in the resulting Feshbach-molecule state $|\text{FB}\rangle$. To measure the effect of light at small detuning from the $X \leftrightarrow b$ transition on the molecules, we illuminate them with a laser beam at a given detuning $\Delta$. This beam is subsequently called the 866-nm beam and is provided by a Ti:sapphire laser locked to a wavelength meter with a systematic frequency error of less than \SI{50}{\mega \hertz}~\cite{Couturier_2018}. This is considered in all frequency errors given in the following. The 866-nm beam is focused to a spot of $1/e^2$-radius 75$\, \mu$m, such that molecules experience an average intensity $I$ of up to $\SI{2700}{W/cm^2}$.

\begin{figure}
\centering
\includegraphics{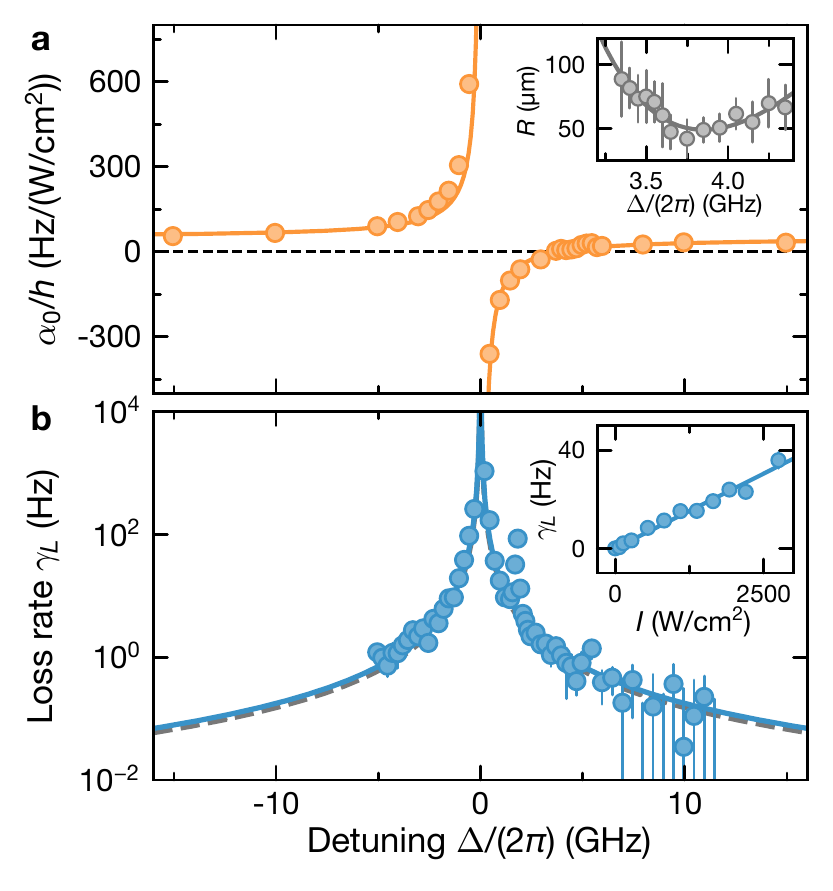}
\caption{Polarizability and loss rate of molecules in state $| 0\rangle$. (a) Experimental data for polarizability $\alpha_0 (\Delta)$ (orange circles) and theoretical curve determined using parameters from intensity-independent measurements (orange line.) The black dashed line indicates zero. Inset: Determination of the tune-out detuning for $| 0 \rangle$ by measuring cloud size after resonant heating. Grey circles are root-mean-square cloud sizes and the solid line is a fit used to find the minimum, see~\cite{supplement}. (b)~Observed loss rate $\gamma_L$ of molecules in $|0\rangle$ subjected to 866-nm light at an intensity of $\SI{1150}{W/cm^2}$ (blue circles). The loss rate at $I=0$ was subtracted from these data points. The blue solid line is a fit of Eq.\ \ref{scattering-rate} with $\omega_0$ and $\Gamma_e$ as the fit parameters, where data points between $\Delta = 2 \pi \times \SI{1.5}{\giga\hertz}$ and $\Delta = 2 \pi \times \SI{2.5}{\giga\hertz}$ were excluded to avoid biasing the fit. The grey dashed line shows the prediction of the photon scattering rate $\gamma_{\text{sc}}$ assuming $\Gamma_e = 2 \pi \times \SI{11.1}{\kilo\hertz}$. Error bars denote the standard error of the fit. Inset: Intensity dependence of the loss rate at $\Delta = 2 \pi \times \SI{1}{\giga \hertz}$. The solid line is a linear fit to the data.}
\label{fig:fig2}
\end{figure}

To directly measure the frequency-dependent polarizability $\alpha_0(\Delta)$ of molecules in the state $| 0\rangle$, we prepared molecules in the crossed dipole trap. The 866-nm beam was turned on during one of the STIRAP pulses and the resulting shift of the STIRAP two-photon resonance was used to determine $\alpha_0$~\cite{supplement}. As shown in Fig.\ \ref{fig:fig2}(a), the data for $\alpha_0$ agree well with the theory curve given by Eq.\ (\ref{polarizability_0}) with parameters determined from the intensity-independent measurements described later.

The tune-out detuning for the $|0 \rangle$ state was also measured with molecules in the crossed dipole trap. In addition to the trap, the 866-nm beam was turned on and modulated for \SI{160}{\milli\second} with $100\%$ peak-to-peak amplitude at a frequency of \SI{110}{\hertz}, equal to the strongest heating resonance of the dipole trap. After this procedure, we measured the molecule cloud size by determining the root-mean-squared deviation of the density distribution, $R$, after \SI{0.6}{\milli\second} time of flight. At $\alpha_0 (\Delta) = 0$, the heating effect is minimized, so that the smallest cloud size should be observed. With the data shown in the inset of Fig.\ \ref{fig:fig2}(a), the tune-out point was determined to be located at $\Delta_{0}^{|0\rangle}= 2 \pi \times \SI{3.85(8)}{\giga\hertz}$. 

In order to trap molecules in an optical dipole trap with long lifetime, the photon scattering rate must be low. We measured the radiative lifetime by illuminating molecules in state $| 0 \rangle$ with 866-nm light. The molecules were frozen in a far-detuned 3D optical lattice to avoid collisional loss. The molecule loss rate $\gamma_{L}$ caused by the 866-nm beam was determined by fitting an exponential decay curve to the measured molecule numbers, see Fig.\ \ref{fig:fig2}(b). From these data, we  determined the position of the resonance feature at $\omega_0 = 2 \pi \times \SI{346.12358(7)}{\tera\hertz}$. To ensure that this resonance was not shifted by the presence of far off-resonant dipole trap light, we performed additional loss measurements for small values of $\Delta$ with all far-detuned trapping light turned off and found a shift in resonance frequency of less than \SI{20}{\mega\hertz}. A calculation with the optical potential method shows that molecules in the $|b^3 \Pi_0\rangle$ state predominantly decay into states in the $|a^3\Sigma^+\rangle$ manifold~\cite{supplement}. We can therefore assume that every photon scattering event leads to the loss of a molecule, such that $\gamma_L \approx \gamma_{\mathrm{sc}}$. Under this assumption, our experimental data yields a value of $\Gamma_e = \SI{13.0(5)}{\kilo\hertz}$, which is in agreement with the theoretical value of $\Gamma_e = \SI{11.1}{\kilo\hertz}$. We additionally investigated the dependence of $\gamma_{L}$ on the light intensity, see inset of Fig.\ \ref{fig:fig2}(b). The observed linear dependence excludes the presence of two-photon scattering processes in this frequency range. At $\Delta = 2 \pi \times \SI{1.78(5)}{\giga\hertz}$ we observed a second, smaller loss peak which we hypothesize could be a transition to a state in the $|b^3\Pi_{0^-} \rangle$ manifold~\cite{supplement}. Still, at all detunings relevant for rotational-state dependent trapping, we find loss rates low enough that lifetimes of more than \SI{1}{\second} can be achieved in a 866-nm trap with a depth of $k_B \times 1\,\mu$K.

\begin{figure}
\centering
\includegraphics{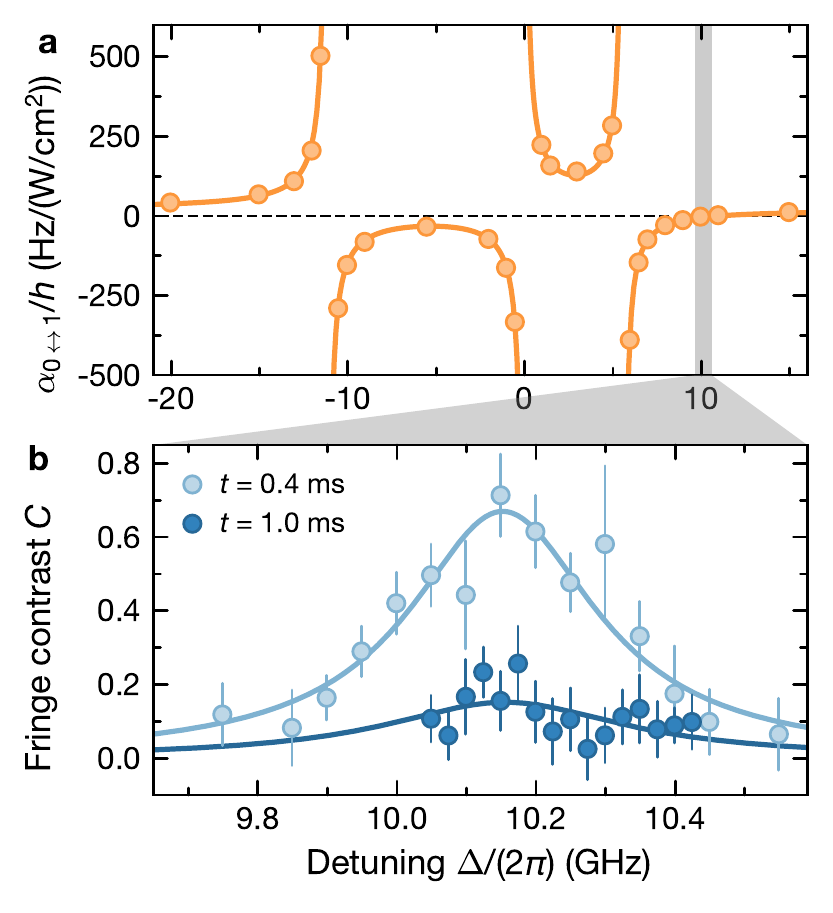}
\caption{Differential polarizability and magic detuning. (a) Experimental data for differential polarizability $\alpha_{0 \leftrightarrow 1}$ from microwave spectroscopy (orange circles) and fit to the data (orange line). The fit function is a combination of three resonances as described by Eq.\ref{polarizability_0}, with a constant offset as well as the linewidths and positions for each resonance as fit parameters. (b) Determination of the magic detuning $\Delta_m$ via Ramsey spectroscopy. Bright (dark) blue circles are experimentally measured contrast after free evolution time $t = \SI{0.4}{\milli\second}(\SI{1.0}{\milli\second})$ in the presence of 866-nm light. Lines are Lorentzian fits to the respective data sets, used to determine the center. Error bars denote one standard deviation and are determined from the covariance matrix of the fits.}
\label{fig:fig3}
\end{figure}

To determine quantities associated with the excited rotational state $|1 \rangle$, we trapped molecules in a spin-decoupled 1D magic lattice described in~\cite{Seesselberg_2018b}. To decouple rotation from nuclear spin and trapping light field, and allow for a well-defined transition to $|1\rangle$, we employed a dc electric field of \SI{86}{V/cm} such that the angle between the polarization of the 866-nm light and the electric field was $4(2)^\circ$. The differential polarizability $\alpha_{0 \leftrightarrow 1} = \alpha_1 - \alpha_0$ was measured via microwave spectroscopy~\cite{supplement}. The resulting data agree with Eqs.\ (\ref{polarizability_0}) and (\ref{polarizability_1}), see Fig.\ \ref{fig:fig3}(a). 

\begin{figure}
\centering
\includegraphics{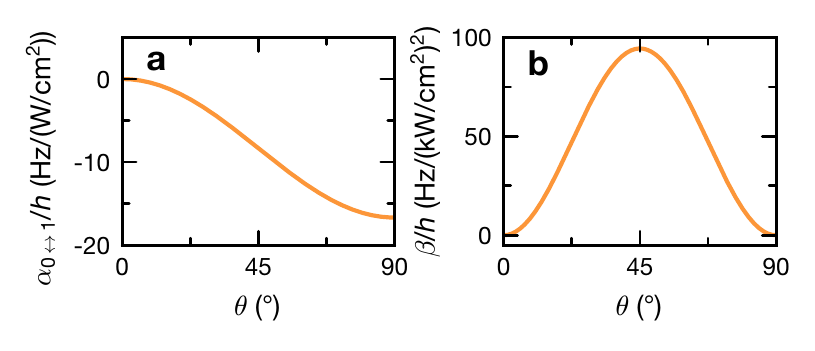}
\caption{Dependence of (a) differential polarizability $\alpha_{0 \leftrightarrow 1}$ and (b) hyperpolarizability $\beta$ in $|1\rangle$ on light polarization angle $\theta$ at the magic detuning with a dc electric field of \SI{86}{V/cm}. $\beta$ represents the second-order dependence of the light shift on the light intensity~\cite{supplement}.}
\label{fig:fig4}
\end{figure}

The magic detuning can be accurately measured via Ramsey spectroscopy of the $|0\rangle \leftrightarrow |1\rangle$ transition, which consists of two resonant $\pi/2$ microwave pulses separated by a free evolution period with duration $t$. We varied the phase $\phi$ of the second microwave pulse for a given $t$ to obtain Ramsey fringes. During the free evolution period, the 866-nm beam was turned on. Any inhomogeneous broadening of the microwave resonance due to the differential light shift of the 866-nm light reduces the contrast of $N_0(\phi)$. Therefore, the magic detuning is identified as the maximum of fringe contrast. By fitting a Lorentzian to the contrast data, see Fig.\ \ref{fig:fig3}(b), we determined the magic detuning to be $\Delta_m = 2 \pi \times \SI{10.15(6)}{\giga\hertz}$. The shift of $\Delta_m$ due to the $4^\circ$ misalignment of $\theta$ is negligible compared to other error sources. Both the first- and second-order differential light shifts vanish and are first-order insensitive to polarization imperfections at $\theta=0^\circ$ as shown in Fig.\ \ref{fig:fig4}. Thus, if $|\theta|\leq0.5^\circ$, the differential light shift of the 866-nm light at magic detuning can be a factor of 70 smaller than that of a typical magic polarization trap. In the present experiments, the coherence time is limited to about \SI{1}{\milli\second} mostly by inhomogeneities of the electric field. Our simulation shows that a coherence time of \SI{30}{\milli\second} in a 866-nm magic trap is feasible if the electric field fluctuation is less than $\SI{0.3}{mV/cm}$~\cite{supplement}. Alternatively, the electric field could be replaced by a strong magnetic field for decoupling the rotation and nuclear spins by Zeeman splitting. The weak coupling between the rotation and magnetic field could allow for longer coherence times.

We can uniquely determine the shape of the polarizability curve $\alpha_0(\Delta)$ from two frequencies that were measured in an intensity-independent manner. The first of these is the tune-out detuning $\Delta_{0}^{|0\rangle}$. The second is the point where the two-photon detuning of STIRAP between the Feshbach-molecule state $|\text{FB}\rangle$ and the state $|0\rangle$ becomes insensitive to the 866-nm light intensity. This is achieved at a detuning $\Delta^\star$ where molecules in $|\text{FB}\rangle$ and $|0\rangle$ experience the same light shift~\cite{supplement, Vexiau_2017}. From these two measured detunings and the value of $\omega_0$, we computed the partial linewidth of the $X \leftrightarrow b$ transition $\Gamma$ as well as the isotropic background polarizability $\alpha_{\text{iso}}$ via Eq.\ (\ref{polarizability_0}). The location of the poles of Eq.\ (\ref{polarizability_1}) and the known ground-state rotational constant $B$ were used to determine the excited-state rotational constant $B^\prime$. To find the values of the background polarizability terms $\alpha_{\text{bg}}^{\parallel}$ and $\alpha_{\text{bg}}^{\perp}$, we used the known form of $\alpha_0(\Delta)$ as well as Eqs.\ (\ref{background_pol_iso})-(\ref{background_pol_ang}) and required the  differential polarizability $\alpha_{0 \leftrightarrow 1}$ to be zero at the measured value of $\Delta_m$. Finally, using $\Gamma$ and the background polarizability terms, we determined the two tune-out detunings of the state $|1 \rangle$ to the left and the right of the $J=1 \leftrightarrow J^\prime=2$ transition, $\Delta_0^{|1\rangle, l}$ and  $\Delta_{0}^{|1\rangle,r}$. In combination, these quantities, summarized in Table \ref{all-quantities}, fully describe the behavior of molecules in the presence of light near the $X \leftrightarrow b$ transition.

\begin{table}
\caption{Summary of the molecular response at the $X \leftrightarrow b$ transition. Our values of $\omega_0$ and $B'$ are compared to literature values for the $^{23}$Na$^{39}$K molecule.}
\begin{ruledtabular}
\begin{tabular}{l l l}
Quantity & Value & Reference\\
\hline
$\omega_0$ & $2\pi\times \SI{346.12358(7)}{\tera\hertz}$ & This work\\
{} & $2\pi\times \SI{346.1434}{\tera\hertz}$ & \cite{Harker_2015} (for $^{23}$Na$^{39}$K)\\
$\Gamma$ & $2\pi\times \SI{301(10)}{\hertz}$ & This work\\
$\Gamma_e$ & $2\pi\times \SI{13.0(5)}{\kilo\hertz}$ & This work\\
$\alpha_{\text{bg}}^\parallel$ & $h \times \SI{105(3)}{Hz/(W/cm^2)}$ & This work\\
$\alpha_{\text{bg}}^\perp$ & $h \times \SI{20(1)}{Hz/(W/cm^2)}$ & This work\\
$B^\prime$ & $h \times \SI{2.79(2)}{\giga\hertz}$ & This work\\
{} & $h \times \SI{2.85}{\giga\hertz}$ & \cite{Harker_2015} (for $^{23}$Na$^{39}$K)\\
$B$ & $h \times \SI{2.8217297(10)}{\giga\hertz}$ & \cite{Will_2016} \\
$\Delta_{0}^{|0\rangle}$ & $ 2 \pi \times \SI{3.85(8)}{\giga\hertz}$ & This work\\
$\Delta_{0}^{|1\rangle,l}$ & $ -2 \pi \times \SI{8.93(15)}{\giga\hertz}$ & This work\\
$\Delta_{0}^{|1\rangle,r}$ & $ 2 \pi \times \SI{7.95(15)}{\giga\hertz}$ & This work\\
$\Delta_m$ & $2 \pi \times \SI{10.15(6)}{\giga\hertz}$ & This work\\
$\Delta^\star$ & $-2 \pi \times \SI{6.78(17)}{\giga\hertz}$ & This work\\
\end{tabular}
\end{ruledtabular}
\label{all-quantities}
\end{table}

In conclusion, we have demonstrated a versatile rotational-state dependent optical dipole trap by utilizing a nominally forbidden electronic transition from the singlet ground state to the lowest electronically excited triplet state of $^{23}$Na$^{40}$K molecules. We precisely determined a tune-out frequency for the ground-state molecules by resonant modulation heating spectroscopy and a magic frequency of rotational states by Ramsey interferometry. Our results open new ways to address key challenges in the field of ultracold polar molecules: for example, trapping molecules in a lattice at one of the tune-out wavelengths would allow selective transfer of hotter molecules at the edge of the lattice into the nontrapped state, thus removing entropy from the sample. In such a lattice, the molecules could thermalize via long-range interactions and would be protected from collisional loss by Pauli blocking~\cite{Chotia_2012} or dipole blocking~\cite{Micheli_2007}, thereby enabling evaporative cooling. The close proximity of the tune-out and magic frequencies allows dynamic switching or even continuous modulation between trap configurations with arbitrary ratios of polarizability experienced by different rotational states. This may open up new possibilities for Floquet engineering of topological states in dipolar spin systems~\cite{Schuster_2019} or other novel methods of dynamic control of quantum systems. Another natural application of light near-resonant to the $X \leftrightarrow b$ transition is to create repulsive potentials for ultracold molecules, e.g.\ to trap them in the dark. Due to the low photon scattering rates at small positive detuning, one can generate a repulsive box trap with sufficiently low intensity in its center to allow investigation of the proposed photon-assisted loss of scattering complexes of molecules~\cite{Christianen_2019a, Gregory_2020, Liu_2020}. Our methods can be generalized to other species of ultracold bialkali molecules by carefully choosing a similar narrow molecular transition.

\begin{acknowledgments}

We thank Y. Bao, H. Bekker, A. Christianen, O. Dulieu, J. He, T. Shi, D. Wang, and T. Zelevinsky for stimulating discussions. We thank C. Gohle, F. See{\ss}elberg, and S. Eustice for their contributions to the experiment. The MPQ team gratefully acknowledges support from the Max Planck Society, the European Union (PASQuanS Grant No. 817482) and the Deutsche Forschungsgemeinschaft under Germany's Excellence Strategy – EXC-2111 – 390814868 and under Grant No. FOR 2247. Work at Temple University is supported by the Army Research Office Grant No. W911NF-17-1-0563, the U.S. Air Force Office of Scientific Research Grant No. FA9550-14-1-0321 and the NSF Grant No. PHY-1908634.

\end{acknowledgments}

\bibliography{bibliography}

\clearpage

\setcounter{figure}{0}
\setcounter{equation}{0}

\renewcommand\thefigure{S\arabic{figure}}
\renewcommand\theequation{S\arabic{equation}}

\newcommand{\tj}[6]{ \begin{pmatrix}
		#1 & #2 & #3 \\
		#4 & #5 & #6 
\end{pmatrix}}

\global\csname @topnum\endcsname 0
\global\csname @botnum\endcsname 0

\section*{Supplemental Material}

\subsection{Theoretical description of polarizability and loss rate}
The matrix elements of the effective molecular Hamiltonian for detuned 
laser-molecule interaction can be written as
\begin{equation}
	\langle i|H_\mathrm{dip}|i\rangle = 
	\sum_{j} \frac{\hbar|\Omega_{ji}|^2}{4\Delta_{ji}} \;,
\end{equation}
using second-order perturbation theory while neglecting the counter-rotating terms.
Here, $\Omega_{ji} = \langle j|\hat{\bm{d}}\cdot\bm{\epsilon}|i\rangle E_0/\hbar$ 
is the Rabi frequency of an electric field with amplitude $E_0$ and 
polarization $\bm{\epsilon}$ on a dipole-allowed transition 
between states $|i\rangle$ and $|j\rangle$, 
$\hat{\bm{d}} = \hat{d}\sum_q \sqrt{\frac{4\pi}{3}}Y_{1,q}\bm{e}^q$ 
is the dipole-moment operator with $\hat{d}$ the corresponding one in the
molecular frame, $Y_{l,m}$ are spherical harmonics, and
$\bm{e}^q$ are the spherical basis unit vectors where $q=0,\pm1$. 
The laser detuning $\Delta_{ji}$ is defined as $(E_j-E_i)/\hbar-\omega$ 
where $E_j$ and $E_i$ are the unperturbed energies of state $|j\rangle$ and
$|i\rangle$ respectively, and $\omega$ is the laser photon frequency.

Now we suppose $|i\rangle$ is one of the two states of interest in the 
vibrational ground state. When $|i\rangle$ is effectively decoupled from
any other energetically nearby states, as realized in the spin-decoupled
setup~\cite{Seesselberg_2018b}, we can approximate its polarizability as
\begin{equation}\label{eq.pol}
	\alpha_i \approx \frac{-\langle i|H_\mathrm{dip}|i\rangle}{I} = -\sum_{j}\frac{z_{ji}^2}{2\hbar\epsilon_0 c}\frac{1}{\Delta_{ji}} \;,
\end{equation}
with the laser field intensity $I$ and $z_{ji} = |\langle j|\hat{\bm{d}}\cdot\bm{\epsilon}|i\rangle|$.

In Born-Oppenheimer approximation, the wavefunction of $|i\rangle$ 
can be written as $|i\rangle = |X^1\Sigma^+\rangle|v=0\rangle|J, m_J\rangle$. 
We use the standard convention where an adiabatic molecular state is described by $^{(2S+1)}\Lambda^{\pm}_\Omega$ with the quantum numbers $S$ as spin, $\Lambda$ 
as the projection of orbital electronic angular momentum onto the molecular axis, 
and $\Omega$ as the projection of the total electronic angular momentum onto 
the molecular axis. 

To study the laser-frequency dependence of the polarizability near the 
narrow $X \leftrightarrow b$ transition, we focus on the contribution 
from the excited states of the $X \leftrightarrow b$ transition, 
which we call $|j^\prime\rangle$ in the summation in Eq.\ \eqref{eq.pol}. 
The contribution from the rest of the states can be approximated to be constant
with respect to laser frequencies within tens of GHz of detuning from
the central frequency $\omega_0$, and is encapsulated in background terms. 
The wavefunction of $|j^\prime \rangle$ can be written similarly as
$|j^\prime\rangle = |b^3\Pi_0\rangle|v^\prime=0\rangle|J^\prime, m_J^\prime\rangle$,
here $|b^3\Pi_0\rangle$ is the electronic wavefunction which mainly has 
$^3\Pi$ character, but also some admixture of $^1\Sigma^+$ character as a result of spin-orbit coupling. We can write it as
$|b^3\Pi_0\rangle = c_1 |b^3\Pi_0\rangle_{\text{diab}} + c_2|A^1\Sigma^+\rangle_{\text{diab}}$, where ``diab'' denotes diabatic potentials before spin-orbit coupling.
It is such mixing that gives the non-zero dipole 
matrix element between the ground states and $|j'\rangle$.

The $q$ part of the matrix element of the dipole-moment operator 
between $|i\rangle$ and $|j^\prime\rangle$ can be written as \cite{Krems_2018}
\begin{eqnarray}
	\nonumber
	&\sqrt{\frac{4\pi}{3}}\bm{e}^q\langle b^3\Pi_0, v', J', m_J'| \hat{d}Y_{1,q} |X^1\Sigma^+, v, J, m_J \rangle =  \\ \nonumber
	&\sqrt{\frac{4\pi}{3}}\bm{e}^q \langle b^3\Pi_0, v' |\hat{d}|X^1\Sigma^+, v \rangle \sum_p  \sqrt{(2J+1)(2J'+1)}\\ 
	&\times \tj{J'}{1}{J}{-m_J'}{q}{m_J}\tj{J'}{1}{J}{-\Lambda'}{p}{\Lambda} \;,
\end{eqnarray}
where $\langle b^3\Pi_0, v' |\hat{d}|X^1\Sigma^+, v \rangle=c_2\langle A^1\Sigma^+_{\text{diab}},v' |\hat{d}|X^1\Sigma^+,v\rangle$ is the Franck-Condon overlap, $p$ is the projection of the scattered photon's angular momentum onto the molecular axis, and $S$, $\Sigma$, and $\Lambda$ are the angular momentum quantum numbers corresponding to the ground and excited electronic states.
Since $\Lambda = \Lambda' = 0$ for $X^1\Sigma^+$ and $A^1\Sigma^+_{\text{diab}}$, 
we have
\begin{eqnarray}\label{eq.me}
\nonumber
&\langle b^3\Pi_0, v', J', m_J' | \hat{d}Y_{1,q} |X^1\Sigma^+, v, J, m_J\rangle = \\ 
\nonumber
&d_0 \sqrt{(2J'+1)(2J+1)} \\
&\times\tj{J'}{1}{J}{-m_J'}{q}{m_J}\tj{J'}{1}{J}{0}{0}{0} \;,
\end{eqnarray}
where $d_0 = c_2\langle A^1\Sigma^+_{\text{diab}},v'=0|\hat{d}|X^1\Sigma^+,v=0\rangle$. 

For state $|0\rangle$, the main contribution to the frequency-dependent part of
the polarizability comes from $|b^3\Pi_0,v' = 0,J' = 1\rangle$, and 
contributions of all other excited 
states can be approximated with a constant $\alpha_\mathrm{iso}$,
defined in the main text. 
From Eq. \eqref{eq.pol} and \eqref{eq.me}, we have 
\begin{eqnarray}\label{eq.pol0}
	\alpha_0 &=& -\frac{2\pi d_0^2}{9 \hbar \epsilon_0 c}\frac{1}{\Delta} + \alpha_\mathrm{iso} 
	= -\frac{3 \pi c^2}{2 \omega_0^3} \frac{\Gamma}{\Delta} + \alpha_{\mathrm{iso}},
\end{eqnarray}
where we introduce the partial linewidth of the transition 
\begin{equation}
\Gamma = \frac{\omega_0^3}{3\pi \epsilon_0 \hbar c^3} z_{ji}^2 = 
\frac{4 \omega_0^3d_0^2}{27\epsilon_0\hbar c^3}.
\end{equation}

For state $|1\rangle$, the frequency-dependent part of the polarizability
comes from the states $|b^3\Pi_0,v' = 0,J' = 0, 2\rangle$, and the frequency-independent part is given by
$\alpha_{\mathrm{iso}}+\alpha_{\text{ang}}(\theta)$, as defined in the main text. 
Unlike for the $|0\rangle$ ground state, the polarizability of $|1\rangle$ depends on the light polarization. For light that is linearly polarized parallel to the quantization axis we have $ \bm{\epsilon}_z =  \bm{e}^{0}$, and for polarization perpendicular to the quantization axis we have 
$ \bm{\epsilon}_{x} = (\bm{e}^{1} + \bm{e}^{-1})/\sqrt{2}$.
With a polarization angle $\theta$, the polarization unit vector can be
written as $\bm{\epsilon}=\bm{\epsilon}_z\cos{\theta}+
\bm{\epsilon}_{x}\sin{\theta}$.
Along with Eq. \eqref{eq.pol} and \eqref{eq.me}, we arrive at the
equation for $\alpha_1$ in the main text.

The photon scattering rate of state $|i\rangle$ is given by 
\begin{equation}
	\gamma_{i} =\sum_{j} \frac{\Omega_{ji}^2}{4\Delta_{ji}^2}\Gamma_j,
\end{equation}
where $\Gamma_j$ is the natural linewidth of $|j\rangle$. For molecules in state $|0\rangle$ interacting with light near the $X \leftrightarrow b$ transition, we have 
\begin{equation}
	\gamma_{sc} = \frac{3 \pi c^2}{2 \hbar \omega_0^3} \frac{\Gamma\Gamma_e}{\Delta^2}I + c^{\text{bg}}I,
\end{equation}
where $\Gamma_e$ is the natural linewidth of $|b^3\Pi_0, v'=0, J'=1\rangle$, and $c^{\text{bg}}I$ includes the contribution from all other excited states which can be neglected for sufficiently small detunings. The excited state $|b^3\Pi_0, v'=0, J'=1\rangle$ can decay to bound or continuum states in the $|X^1\Sigma^+\rangle$ or $|a^3\Sigma^+\rangle$ electronic potentials. The total contribution to the natural linewidth from states in each potential can be estimated using the optical potential approach~\cite{Zygelman_1988}. This yields
\begin{equation}
 \Gamma_{S} = \int_0^\infty \mathrm{d}R \,
   \phi_e(R)^\ast\phi_e(R)A_S(R) \;,
\end{equation}
where the final state $|S\rangle$ can be $|X^1\Sigma^+\rangle$ or $|a^3\Sigma^+\rangle$, $R$ is the internuclear separation, and $\phi_e(R)$ is the vibrational wavefunction of $|b^3\Pi_0, v'=0, J'=1\rangle$. $A_S(R)$ is the optical potential, given by
\begin{equation}
  A_S(R) = \frac{1}{3 \pi \epsilon_0 \hbar^4} D_{eS}^2(R)
   \frac{|E_e(R)-E_S(R)|^3}{c^3} \;,
\end{equation}
where $D_{eS}(R)$ is the transition dipole moment
between the states $|b^3\Pi_{0}\rangle$ and $|S\rangle$, $E_e(R)$ and $E_S(R)$ are the potential energies of these states, respectively, and $c$ is
the speed of light. We construct the potential energies of $|b^3\Pi_{0}\rangle$ from diabatic potentials and spin-orbit functions given in Ref.\ \cite{Harker_2015}.
The $|X^1\Sigma^+\rangle$ and $|a^3\Sigma^+\rangle$
potentials are taken from Ref.\ \cite{Gerdes_2008}.
The non-relativistic transition dipole moments are 
taken from Ref.~\cite{Aymar_2007}. From this, we find that
the contribution to $\Gamma_e$ from states in the $|X^1\Sigma^+\rangle$ 
potential is $2\pi \times \SI{1.0}{\kilo\hertz}$, and the contribution from
states in the $|a^3\Sigma^+\rangle$ potential is
$2\pi \times \SI{10.1}{\kilo\hertz}$, which is nearly four orders of magnitude larger than in the case of KRb. The reason is that the transition dipole moment of the
$|b^3\Pi_{0}\rangle \leftrightarrow |a^3\Sigma^+\rangle$ transition at the bottom of the $|b^3\Pi_{0}\rangle$ potential is $d = 0.48 \, ea_0$ for NaK, while $d = 0.006(4)\, ea_0$ for KRb, with the elementary charge $e$ and the Bohr radius $a_0$. Decay into the $|X^1 \Sigma^+, v=0, J=2 \rangle$ state is expected to be twice as large as into the $J=0$ ground state, while decay into vibrational states other than $v=0$ is small. $\Gamma$ then equals one third of the contribution from states in the $|X^1\Sigma^+\rangle$  potential, which agrees with our experimental value of $\Gamma = 2 \pi \times \SI{301(10)}{\hertz}$. Thus the branching ratio
$\Gamma/\Gamma_e$ is $0.03$. 

The differential light shift between the $|0\rangle$ and $|1\rangle$ states
can be approximated by \cite{Seesselberg_2018b} 
\begin{equation}
	\delta \omega_{0 \leftrightarrow 1} = \frac{1}{\hbar}(\alpha_{0 \leftrightarrow 1}(\theta)I+\beta(E,\theta)I^2+\mathcal{O}(I^3)),
\end{equation}
where $\theta$ is the polarization angle, $\alpha_{0 \leftrightarrow 1}$ 
is the differential polarizability as defined in the main text,
$\beta$ is the hyperpolarizability of $|1\rangle$, and $E$ is the magnitude of the applied dc electric field. 
An approximation for $\beta$ can be derived by considering
the contribution from four-photon couplings to the
$|X^1\Sigma^+,v = 0,J = 1, m_J=\pm 1\rangle$ states and back. It reads
\begin{equation}
	\beta(E,\theta) = \frac{5B}{3d^2E^2}(\alpha_1(0)-\alpha_1(\pi/2))^2\sin^2(2\theta),
\end{equation}
where $d = \SI{2.72}{D}$ is the permanent dipole moment of $^{23}$Na$^{40}$K.

\subsection{Experimental setup}
Molecule association is performed after preparing a mixture of about $10^5$ $^{23}$Na and $^{40}$K atoms each, at a temperature of \SI{300}{\nano\kelvin} and a magnetic field of \SI{85.4}{G} in the vertical ($z$) direction. We then apply a radiofrequency pulse to create molecules in a weakly bound Feshbach-molecule state $|\text{FB}\rangle$ and use STIRAP as described in~\cite{Seesselberg_2018a} to create molecules in the rovibrational ground state $|0\rangle$. After STIRAP, we use short pulses of resonant light to remove remaining unassociated sodium and potassium atoms. The association procedure can be done either in a far-detuned crossed-beam optical dipole trap or in a 1D or 3D optical lattice. The crossed dipole trap consists of a 1064-nm and a 1550-nm laser beam intersecting orthogonally in the horizontal ($x$-$y$) plane. The trap frequencies experienced by molecules in $|0\rangle$ in this trap are (94, 72, 233) Hz in the ($x, y, z$)-directions, respectively. The 1D lattice is formed by a retro-reflected 1550-nm laser beam and is magic for the $|0\rangle \leftrightarrow |1\rangle$ transition. As described in~\cite{Seesselberg_2018b}, this is achieved by applying a dc electric field in the $y$-direction, which serves to decouple the rotational states, the hyperfine states, and the trapping light field, and by aligning the polarization of the lattice light with a magic angle relative to this electric field. The 3D lattice is used to suppress collisional loss in experiments that require long molecule lifetimes. It is formed by three retro-reflected laser beams: In the vertical direction, the wavelength is \SI{1550}{\nano \meter} and the beam size is $100\,\mu$m, allowing for lattice depths of up to $800\,E_R$ for ground-state molecules, where $E_R$ is the photon-recoil energy for these molecules in a lattice of the respective wavelength. In both horizontal directions, the wavelength is \SI{1064}{\nano \meter} and the beam size is $300\,\mu$m. The maximal lattice depth in these directions is $200\,E_R$. The 866-nm beam is focused onto the molecules along the $z$-direction. Except in the measurements to determine the polarization dependence of the differential polarizability, the polarization of this beam is always at an angle of $4(2)^\circ$ to the $y$-direction, almost parallel to the dc electric field.

\subsection{Tune-out detuning determination}
The tune-out detuning $\Delta_{0}^{|0\rangle}$ for molecules in state $|0\rangle$ was identified as the detuning where the minimum of heating occurs when modulating the 866-nm beam at a heating resonance of the crossed dipole trap. The heating process depends on the sample temperature as well as the modulation amplitude and modulation frequency. When the temperature of the molecules is much smaller than the trap depth and the modulation is weak, the effect of heating can be described as a linear increase in the sample's energy at a rate $S \alpha^2_0(\Delta) I_{\text{mod}}^2$, where $S$ depends on the modulation frequency and initial temperature and $I_{\text{mod}}$ is the intensity modulation amplitude~\cite{Gehm_1998}. For the case of strong heating, the temperature quickly saturates to an equilibrium where the heating is balanced by hot molecules escaping from the trap. However, for $\alpha_0(\Delta) \approx 0$, the linear model is still valid. The expression for the cloud size $R$ after modulating the 866-nm beam power at a given modulation frequency for a fixed time then reads
\begin{equation}
\label{tuneoutfit}
R(\Delta) = R_0 + \chi \left( \frac{1}{\Delta} - \frac{1}{\Delta_{0}^{|0\rangle}} \right)^2.
\end{equation}
Here, $R_0$ is the initial cloud size and $\chi$ is a constant which contains the dependence on intensity, modulation time and modulation frequency. We used this expression with $R_0$, $\chi$, and $\Delta_{0}^{|0\rangle}$ as fit parameters to determine the detuning at which the minimum of heating occurs and thereby find the tune-out detuning, as shown in the inset of Fig. 2(a) of the main text.

\subsection{Magic condition between Feshbach and ground-state molecules}

\begin{figure}
\centering
\includegraphics{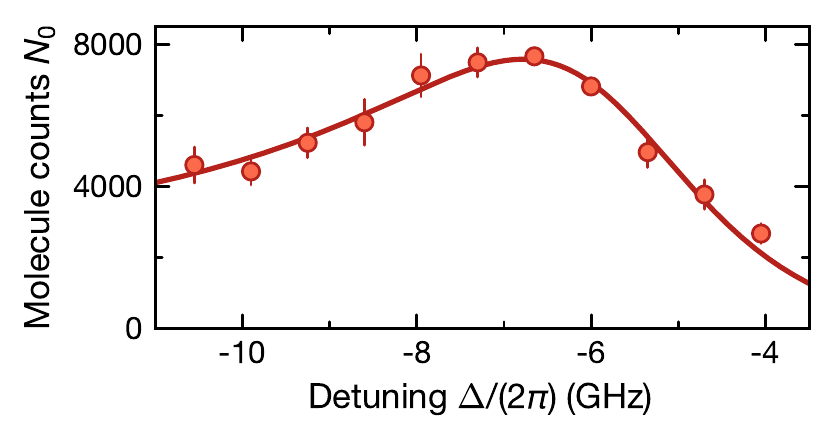}
\caption{Magic condition for $|\text{FB}\rangle$ and $|0\rangle$. The STIRAP two-photon detuning was set to the resonance frequency as determined with no 866-nm light. Data points were taken with 866-nm light at various $\Delta$ present during STIRAP (red circles). The solid line is a fit of Eq.\ (\ref{fbfit}). Error bars denote the standard error of the mean of 6 data points.}
\label{fig:figS1}
\end{figure}

We identified the magic detuning $\Delta^{\star}$ from $\omega_0$ where molecules in the Feshbach-molecule state $|\text{FB} \rangle$ and the rovibrational ground state $|0 \rangle$ experience the same light shift in a way that is independent of the intensity of the 866-nm light. To do this, 866-nm light at various values of $\Delta$ was turned on during one of the STIRAP pulses at a two-photon detuning which was calibrated to be resonant in the case with no 866-nm light. The individual light shifts of $|\text{FB}\rangle$ and $|0\rangle$ detune the STIRAP two-photon resonance and thereby lower the molecule-conversion efficiency unless $\alpha_0(\Delta)$ matches the polarizability $\alpha_{\text{FB}}$ of the molecules in $|\text{FB} \rangle$ at $\Delta^{\star}$. Because $| \text{FB} \rangle$ is a very weakly bound state, its polarizability can be computed to be $\alpha_{\text{FB}} = h \times \SI{76.26}{Hz/(W/cm^2)}$ by summing the polarizabilities of the constituent atoms~\cite{Vexiau_2017, Grimm_2000, Safronova_2006}. This number is approximately independent of $\Delta$ because the molecular resonance is far below the lowest atomic resonances of $^{23}$Na and $^{40}$K. To determine $\Delta^\star$ from the data, we model the drop in STIRAP conversion efficiency due to the shift of the STIRAP two-photon resonance as
\begin{equation}
\label{fbfit}
N_0 = N_{\text{max}} \frac{\Gamma_s^2/4}{\Gamma_s^2/4 + (I (\alpha_0(\Delta) - \alpha_{\text{FB}})/\hbar)^2},
\end{equation}
where $N_0$ is the number of molecules that we detect in state $|0\rangle$, $N_{\text{max}}$ is the number of detected molecules when the STIRAP two-photon transition is on resonance, and $\Gamma_s$ is the linewidth of the STIRAP two-photon resonance. To obtain a fit function, the general form $\alpha_0(\Delta) = A/\Delta + \alpha_c$ was inserted into Eq.\ (\ref{fbfit}), and $N_{\text{max}}, \Gamma_s, A$, and $\alpha_c$ were used as fit parameters. We found the maximum conversion efficiency at $\Delta^\star = - 2 \pi \times \SI{6.78(17)}{\giga \hertz}$, see Fig.\ \ref{fig:figS1}. The values found for $A$ and $\alpha_c$ are consistent with the parameters given in Table 1 of the main text within their error bars.

\subsection{Intensity calibration}
\begin{figure}
\centering
\includegraphics{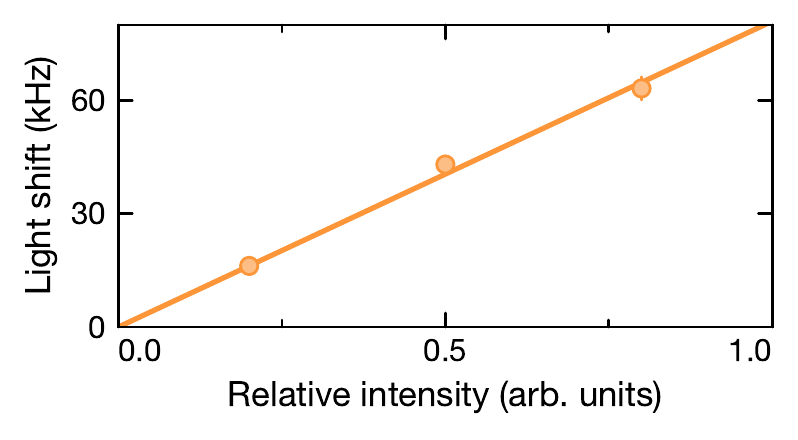}
\caption{Intensity calibration of the 866-nm light from known polarizability at $\Delta = 2 \pi \times \SI{80}{\giga\hertz}$. Differential light shift between $|0\rangle$ and $\text{FB}\rangle$ measured via STIRAP two-photon resonance shift $\delta \omega_{\text{FB} \leftrightarrow 0}$ for different intensities. Circles denote the center frequencies of Lorentzian fits to the spectra recorded at each intensity, error bars are derived from the covariance matrix of the fit. The line is a linear fit to the center frequencies.}
\label{fig:figS2}
\end{figure}

The intensity of the 866-nm light was calibrated from the measured light shift and the known polarizabilities of ground-state molecules $\alpha_0$ and of Feshbach molecules $\alpha_{\text{FB}}$ at $\Delta = 2 \pi \times \SI{80}{\giga\hertz}$ according to
\begin{equation}
\hbar \delta \omega_{\text{FB}\leftrightarrow 0}(I,\Delta) = (\alpha_0(\Delta)-\alpha_{\text{FB}}) I
\end{equation}
as shown in Fig.\ \ref{fig:figS2}. Specifically, we used the value $\alpha_0(2 \pi \times \SI{80}{\giga\hertz}) = h \times \SI{47(1)}{Hz/(W/cm^2)}$, which was obtained from the previous measurements of $\Delta_0^{|0\rangle}$ and $\Delta^\star$. We found 2700(100)\,W/cm$^2$ at 100\% relative power.

\subsection{Polarizability measurements}
\begin{figure}
\centering
\includegraphics{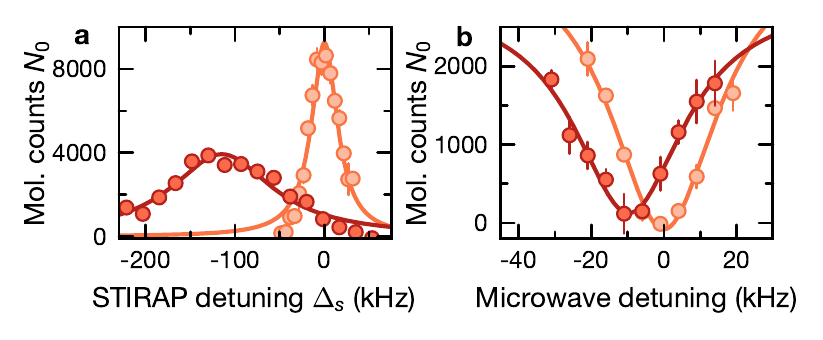}
\caption{Example data for polarizability measurements. (a)~Determination of $\alpha_0$ via STIRAP two-photon resonance shift. The 866-nm beam was turned on during STIRAP. Data were taken at $\Delta = -2 \pi \times \SI{2}{\giga \hertz}$ and $I = \SI{1200}{W/cm^2}$ (dark red) and compared to a calibration measurement at $I=0$ (bright red).
(b) Determination of differential polarizability $\alpha_{0 \leftrightarrow 1}$ via microwave spectroscopy. Data were taken at $\Delta = 2 \pi \times \SI{3}{\giga \hertz}$ and $I = \SI{69}{W/cm^2}$ (dark red) and compared to a calibration measurement at $I=0$ (bright red). The solid lines are Lorentzian fits. Error bars denote the standard error of the mean of 3 to 4 data points.}
\label{fig:figS3}
\end{figure}

The polarizability $\alpha_0(\Delta)$ of molecules in state $| 0 \rangle$ was determined from the observed shift of STIRAP two-photon resonance that occurred when turning the 866-nm beam on during one of the STIRAP pulses. Example data is shown in Fig.\ \ref{fig:figS3}(a). The shift of two-photon detuning is equal to the differential light shift $\hbar \delta \omega_{\text{FB} \leftrightarrow 0}(I, \Delta)$ between the $| \text{FB} \rangle$ and  $| 0\rangle$ states. From this, we obtained $\alpha_0(\Delta)$ via

\begin{equation}
\alpha_0 (\Delta) =  \alpha_{\text{FB}} - \hbar \delta \omega_{\text{FB} \leftrightarrow 0}(I, \Delta) / I.
\end{equation}
The precision of this method is limited by drifts of the STIRAP laser frequency, which we compensated as far as possible by performing regular calibration measurements without 866-nm light.

The differential polarizability $\alpha_{0 \leftrightarrow 1}(\Delta)$ was measured via microwave spectroscopy. After the association of molecules in the state $| 0 \rangle$ in the magic 1D lattice, their rotational state can be changed to $| 1 \rangle$ via a resonant microwave $\pi$-pulse with a duration of 35\,$\mu$s. This can be observed as molecule loss because molecules in $| 1 \rangle$ are not resonant with the reverse STIRAP. The light shift $\hbar \delta \omega_{0 \leftrightarrow 1}(\Delta)$ of the $|0\rangle \leftrightarrow |1\rangle$ transition caused by the presence of 866-nm light during the microwave pulse then yields $\alpha_{0 \leftrightarrow 1}(\Delta)$ by
\begin{equation}
\alpha_{0 \leftrightarrow 1}(\Delta) \equiv \alpha_1 (\Delta) - \alpha_0 (\Delta) = \hbar \delta \omega_{0 \leftrightarrow 1}(\Delta) / I.
\end{equation}
Example data from a scan of the microwave transition frequency is shown in Fig.\ \ref{fig:figS3}(b). For all polarizability measurements, the intensity of the 866-nm light was chosen in order to achieve a compromise between the magnitude of the light shift and the inhomogeneous broadening caused by the finite size of the 866-nm beam. For the measurements of $\alpha_0$,  intensities between $\SI{360}{W/cm^2}$ and $\SI{2200}{W/cm^2}$ were used. The measurements of $\alpha_{0 \leftrightarrow 1}$ were performed at intensities between $\SI{69}{W/cm^2}$ and $\SI{550}{W/cm^2}$. 

\subsection{Polarization dependence}
\begin{figure}
\centering
\includegraphics{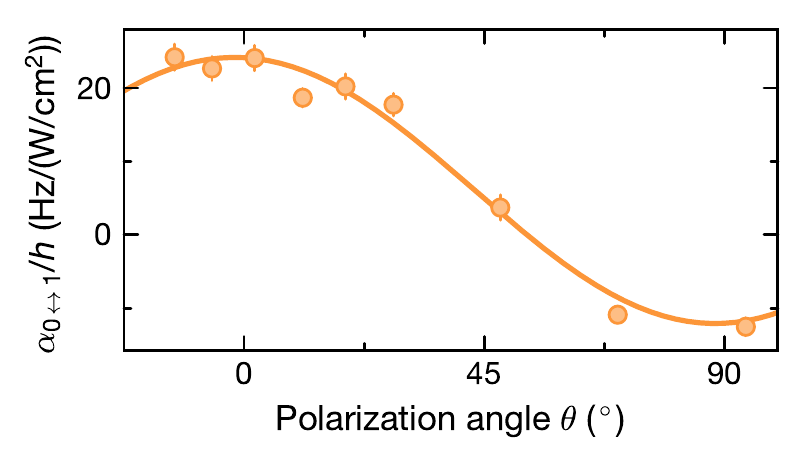}
\caption{Dependence of differential polarizability $\alpha_{0 \leftrightarrow 1}$ on the angle $\theta$ between laser polarization and electric field. Measurements were performed at $\Delta = 2 \pi \times 80$ GHz. Error bars denote the standard error of the mean of 4 to 8 data points.  The solid line is a fit of Eq.\ (4) of the main text.}
\label{fig:figS4}
\end{figure}

The polarization dependence of the polarizability $\alpha_{\text{ang}} (\theta)$ was determined by measurements of the differential polarizability $\alpha_{0 \leftrightarrow 1}$ at a constant detuning  $\Delta = 2 \pi \times \SI{80}{\giga\hertz}$ and at various angles between the laser polarization and the electric field, see Fig.\ \ref{fig:figS4}. At this detuning, $\alpha_{0 \leftrightarrow 1} = \alpha_{\text{ang}} (\theta)$ is a good approximation. The results agree with the prediction of Eq.\ (4) in the main text as well as with the values determined for $\alpha_{\text{bg}}^\parallel$ and $\alpha_{\text{bg}}^\perp$ in Table I of the main text. Another independent determination of $\alpha_{\text{ang}}(0)$ from the offset of the fit shown in Fig.\ 3(a) of the main text also yields consistent results.

\subsection{Ramsey spectroscopy}
\begin{figure}
\centering
\includegraphics{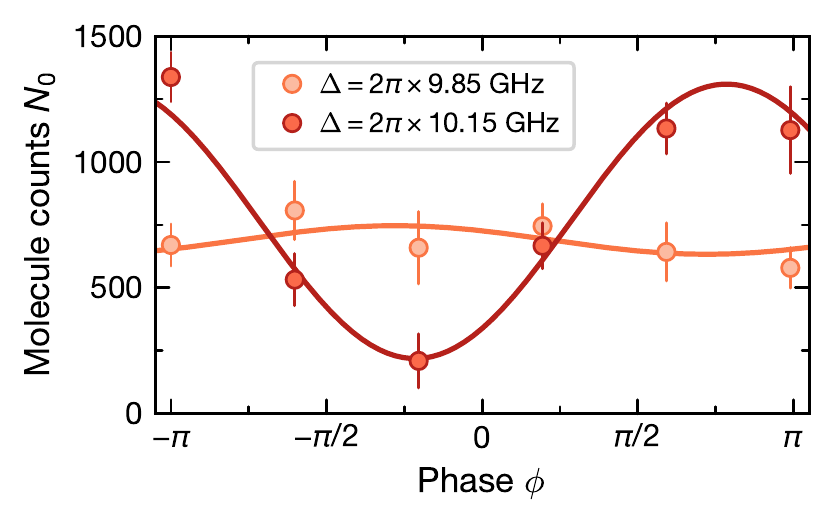}
\caption{Example data for determination of the magic detuning via Ramsey spectroscopy. Data were taken after \SI{0.4}{\milli\second} free evolution time at detunings $\Delta = 2 \pi \times \SI{9.85}{\giga \hertz}$ (bright circles) and $\Delta = 2 \pi \times \SI{10.15}{\giga \hertz}$ (dark circles). Error bars denote the standard error of the mean of 4 data points. The solid lines are fits of Eq.\ (\ref{ramsey-fringe}).}
\label{fig:figS5}
\end{figure}

Ramsey spectropscopy was used to determine the magic frequency, at which the smallest dephasing occurs for superpositions of the states $|0\rangle$ and $|1\rangle$. 866-nm light at a given detuning was turned on during the free evolution time $t$. To mitigate a damped and chirped interference fringe due to the fast drift of the electric field and molecule loss, instead of changing $t$ between measurements, we varied the phase $\phi$ of the second microwave pulse for a given $t$. The fringe contrast $C$ and initial phase $\phi_0$ were determined by fitting the function
\begin{equation}
\label{ramsey-fringe}
N_0(\phi) = \frac{N_{\textrm{tot}}(t)}{2}\left( 1 - C(t)\cos(\phi + \phi_0) \right)
\end{equation}
to the measured molecule numbers, see example data shown in Fig.\ \ref{fig:figS5}.

\begin{table*}
\caption{Estimated dephasing rates at the magic condition with $E=\SI{86}{V/cm}$. The first section contain the dephasing caused by differential light shifts from the 866-nm light. We assume a frequency error $\delta\omega=2\pi\times\SI{10}{MHz}$ and a polarization angle $\theta = 4^\circ$ with a fluctuation of $\delta\theta=0.5^\circ$. The second section contains dephasing caused by the 1550-nm light, assuming a polarization within $0.5^\circ$ of the magic angle. Finally, we consider the short-term noise (rms value) and inhomogeneity of the electric field, denoted $\delta E_{N}$ and $\delta E_{G}$, respectively, and the dipolar interaction between molecules.}
\begin{ruledtabular}
\begin{tabular}{l l l} 
   Source & Value & Dephasing rate\\ 
   \hline
   $\beta$ & $h\times 1.7~\text{Hz/(kW/cm}^2)^2$ & 
   \multirow{3}{*}{
    $\Bigg\} \; h \times 37\,$Hz}\\
	$\delta\alpha_{0 \leftrightarrow 1}$ from angle error & $ h\times 0.02~\text{Hz/(W/cm}^2)$ \\
   $\delta\alpha_{0 \leftrightarrow 1}$ from frequency error & $ h\times 0.06~\text{Hz/(W/cm}^2)$ \\ 
   \hline
   $\beta_{1550}$ & $h\times 19~\text{Hz/(kW/cm}^2)^2$ & 
   \multirow{2}{*}{
   	$\Big\} \; h \times 36\,$Hz}\\ 
   $\delta\alpha_{0 \leftrightarrow 1,1550}$ from angle error & $ h\times 0.07~\text{Hz/(W/cm}^2)$ \\ 
   \hline
   $\delta E_{N}$ & \SI{0.5}{mV/cm} & $ h \times 21\,$Hz\\ 
   $ \delta E_{G}$ & $8$ mV/cm & $ h \times 244\,$Hz \\ 
   \hline
   Dipolar interaction & & $ h \times 20\,$Hz
\end{tabular}
\end{ruledtabular}
\label{Tab1}
\end{table*}

\subsection{Estimation of single-particle dephasing rate}
We use the same method as in \cite{Seesselberg_2018b} to estimate the total dephasing rate $\Gamma_d$, which is defined such that the $T_2$ coherence time is given by $T_2 = 2 \hbar / \Gamma_d$. There are multiple contributions to $\Gamma_d$: $\Gamma_{EG}$ due to spatial variation of the electric field, $\Gamma_{EN}$ due to short-term noise of the electric field, $\Gamma_{L}$ due to the differential light shift of the 866-nm light, and $\Gamma_{L,1550}$ due to the differential light shift of the 1550-nm light. If all the sources are uncorrelated, the total dephasing rate is given by $\Gamma_d\approx\sqrt{\Gamma_{EG}^2+\Gamma_{EN}^2+\Gamma_L^2+\Gamma_{L,1550}^2}$, which is verified by numerical simulation~\cite{Seesselberg_2018b,Neyenhuis_2012,Blackmore_2018}. An overview of the dephasing sources is given in Table \ref{Tab1}. The dominating source of dephasing in our experiments is electric field inhomogeneity. Although it is hard to exactly measure the electric field distribution inside the molecular cloud, a previous calibration yielded electric field gradients on the order of $\SI{1.5}{V/cm^2}$ at a dc-offset field of $E = \SI{68.3}{V/cm}$. This results in an estimate of the electric field inhomogeneity $\delta E_G=\SI{8}{mV/cm}$ over the molecular cloud at $E = \SI{86}{V/cm}$, assuming that there is only a linear inhomogeneity which is proportional to $E$, and a $1/e$ cloud radius of $\SI{30}{\mu m}$ in the horizontal directions and of $\SI{10}{\mu m}$ in the vertical direction. The resulting dephasing rate is $\Gamma_{EG}\approx2\xi E\delta E_G=h\times\SI{244}{Hz}$, where $\xi=h\times \SI{177}{Hz/
(V/cm)^2}$. The next important source of dephasing is the residual differential light shift $\alpha_{0 \leftrightarrow 1}$ caused by laser frequency drifts and errors in the polarization angle $\theta$ in combination with the finite size of the 866-nm beam. From the molecule density profile and the beam shape, we estimate the intensity variation over the molecular cloud to be $\delta I_{866} = \SI{589}{W/cm^2}$, which is consistent with the width of the fitted curve shown in Fig.\ 3(b) of the main text. The resulting dephasing rate is approximately $\Gamma_L=\alpha_{0 \leftrightarrow 1}\delta I_{866}$ since the second-order light shift is negligible. Near the magic detuning $\Delta_m$, the differential polarizability changes linearly with the laser frequency: $\alpha_{0 \leftrightarrow 1}(\Delta)=\SI{0.006}{Hz/(W/cm^2)}\times(\Delta - \Delta_m)/(2 \pi \, \text{MHz})$. The frequency drift of the 866-nm light between experimental cycles is given by the relative precision of the wavelength meter used to lock the laser, which results in a root-mean-square (rms) value of $\delta\omega=2\pi\times\SI{10}{MHz}$ on time scales of hours. The term caused by imperfect laser polarization depends in second-order on the polarization angle: $\alpha_{0 \leftrightarrow 1}(\theta)=\SI{-0.0053}{Hz/(W/cm^2)\times \theta^2/(1^\circ})^2$. The typical rms fluctuation of the laser polarization due to temperature and stress-induced drift of optical components is $\delta\theta=0.5^\circ$. Although the differential light shift due to the $4^\circ$-misalignment of the laser polarization can be mostly compensated by tuning the laser frequency slightly away from the magic detuning for $\theta=0^\circ$, it increases the sensitivity to the angle fluctuation. The resulting total dephasing rate from the 866-nm light is then $\Gamma_L = h\times\SI{37}{Hz}$ assuming uncorrelated frequency and polarization fluctuation. This is comparable to the rate from the spin-decoupled magic trap formed by the 1550-nm light for which we estimate an intensity variation $\delta I_{1550}=\SI{520}{W/cm^2}$ and assume a laser polarization $\theta_{1550}$ aligned to the magic angle with $\delta \theta_{1550} = 0.5^\circ$. Together, these factors give a conservative upper bound of \SI{1.3}{ms} for the coherence time.

\begin{figure}
\centering
\includegraphics{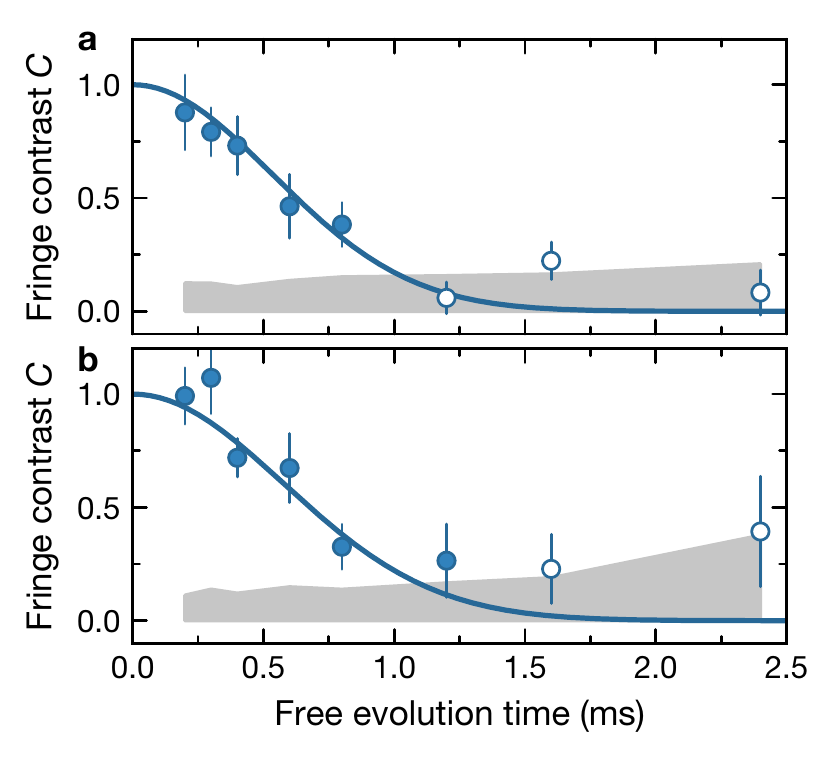}
\caption{Contrast in Ramsey spectroscopy after variable free evolution time. (a) With 866-nm light turned on, (b) with 866-nm light turned off. Blue points denote contrast data, blue lines are Gaussian fits used to determine the $T_2$ coherence time. The grey shaded area indicates the contrast bias caused by molecule number fluctuations. Open circles denote data points that were ignored in the fit due to bias, see \cite{Seesselberg_2018b}.}
\label{fig:figS6}
\end{figure}

To compare this estimation to our experiments, we took data of the Ramsey contrast versus free evolution time, shown in Fig.\ \ref{fig:figS6}. The coherence times extracted from these data in presence and absence of 866-nm light are \SI{0.75(4)}{\milli\second} and \SI{0.81(7)}{\milli\second}, respectively. These values qualitatively agree with the estimated $T_2$ based on independent calibrations, and also show that the dephasing is dominated by factors that are not caused by the 866-nm light. In Ref.~\cite{Seesselberg_2018b}, we realized with a comparable molecule number of 1500 a coherence time of about \SI{5}{\milli\second} in a spin-decoupled magic trap of identical trap geometry. The major difference to this previous experiment is that we did not use additional electrodes to compensate electric field gradients in the measurements shown here. By comparing the coherence times, we deduce an extra dephasing rate of $h\times\SI{388}{Hz}$ which suggests an electric field inhomogeneity of 12.7 mV/cm. The discrepancy to the previous gradient calibration could be caused by drifts of the electric field gradients, electric field curvature, or polarization misalignment of the 1550-nm trap, which were not accounted for in the calculations above.

\begin{table}
\caption{Estimated dephasing rate for the 866-nm light with improved laser frequency stability and laser polarization alignment. We assume $\delta\omega=2\pi\times\SI{100}{kHz}$ and $\theta = 0^\circ$ with a fluctuaton of $\delta\theta=0.5^\circ$.}
\begin{ruledtabular}
\begin{tabular}{l l l} 
   Source & Value & Dephasing rate \\ 
   \hline
   $\beta$ & $h\times 0.03~\text{Hz/(kW/cm}^2)^2$ & \multirow{3}{*}{
   	$\Bigg\} \; h\times\SI{0.9}{Hz} $}\\ 
	$\delta\alpha_{0 \leftrightarrow 1}(\delta\theta)$ & $ h\times 0.0014~\text{Hz/(W/cm}^2)$  \\
   $\delta\alpha_{0 \leftrightarrow 1}(\delta\omega)$ & $ h\times 0.0006~\text{Hz/(W/cm}^2)$ \\ 
\end{tabular}
\end{ruledtabular}
\label{Tab2}
\end{table}

\begin{figure}
\centering
\includegraphics{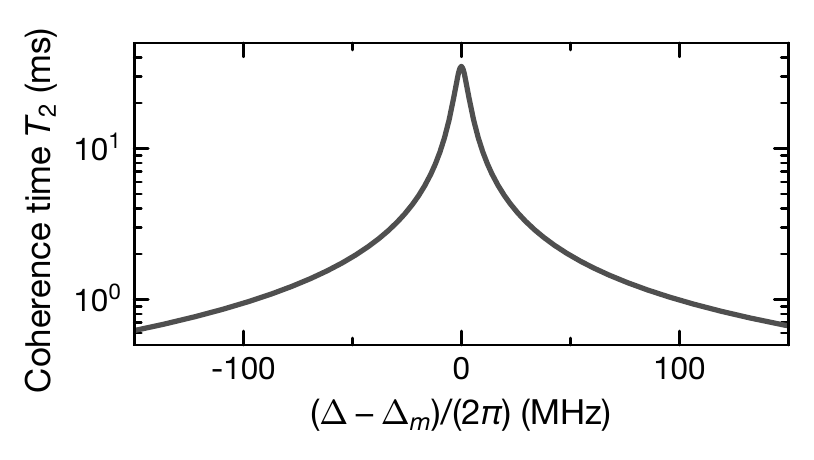}
\caption{Simulation of coherence time in a 866-nm dipole trap as a function of laser detuning for an effective electric-field fluctuation $\delta E = \SI{0.3}{mV/cm}$ at $E=\SI{86}{V/cm}$. We assume $\theta = 0^\circ$ with a fluctuation of $\delta\theta = 0.5^\circ$ and neglect the finite laser linewidth as well as the dipolar interaction between molecules given a very dilute molecule gas.}
\label{fig:figS7}
\end{figure}

Significant improvements of the coherence time should be possible with the 866-nm light. In fact, a dipole trap that consists only of light at the magic wavelength, assuming a laser frequency locked to the magic detuning within $2\pi\times\SI{100}{kHz}$ and polarization aligned to the electric field within $0.5^\circ$, would allow for dephasing rates as low as $h \times \SI{0.9}{Hz}$, see Table \ref{Tab2}. If we normalize to the absolute light shift across the molecular cloud, the differential light shifts per trap depth are $h\times 1\,$Hz/($k_B\times\mu$K) for the 866-nm magic frequency trap and $h\times 73\,$Hz/($k_B\times\mu$K) for the 1550-nm spin-decoupled magic polarization trap, assuming they are both aligned to the respective magic angle with the same fluctuations of $0.5^\circ$. However, under realistic experimental conditions, the electric field inhomogeneity will become the limiting factor. The total single-particle dephasing rate can then be approximated as
\begin{equation}
\label{dephasing-rate}
\Gamma_d \approx \sqrt{(\alpha_{0 \leftrightarrow 1}(\Delta,\theta=0)\delta I_{866})^2+{\Gamma_0}^2},
\end{equation}

where $\Gamma_0=2\xi E \delta E$ and $\delta E=\sqrt{2\delta E_N^2+\delta E_G^2}$ is the effective electric-field fluctuation combining short-term noise $\delta E_N$ and inhomogeneity $\delta E_G$. The $T_2$ coherence time is shown as a function of laser detuning for $\delta E = \SI{0.3}{mV/cm}$ in Fig.\ \ref{fig:figS7}. These results demonstrate that $T_2 = \SI{30}{ms}$ can be reached even with a laser-frequency error of $2\pi\times\SI{3}{MHz}$, which is well within state-of-the-art in cold-atom experiments. 

\subsection{Lifetime measurements}
\begin{figure}
\centering
\includegraphics{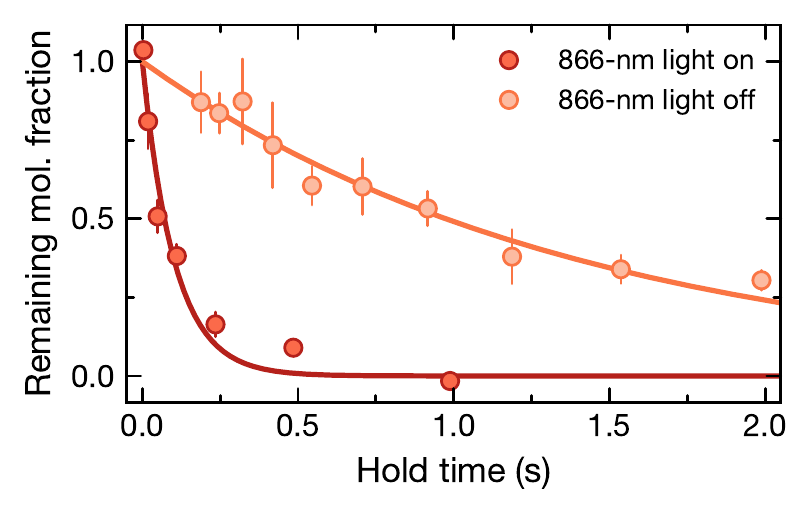}
\caption{Example data for lifetime measurements in a 3D lattice in the presence of 866-nm light at a detuning $\Delta = - 2 \pi \times \SI{1.5}{\giga \hertz}$ and intensity $I = \SI{1150}{W/cm^2}$ are shown in dark red. The measurement of the background loss rate in the deep 3D lattice is shown in bright red. Circles are experimental data, error bars denoting the standard error of the mean of 3 to 4 data points, and the solid lines are exponential fits to determine the $1/e$ lifetime.}
\label{fig:figS8}
\end{figure}

\begin{figure}
\centering
\includegraphics{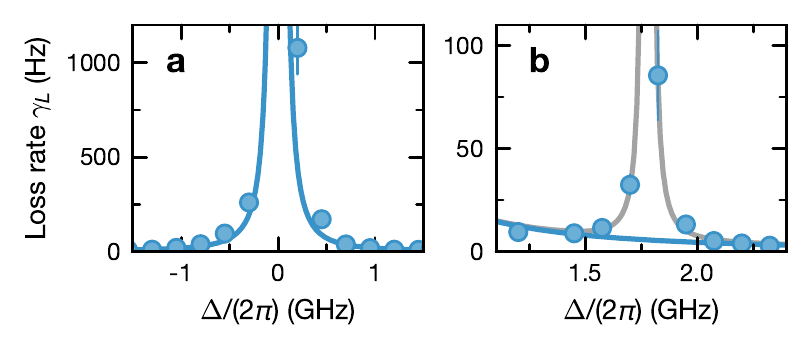}
\caption{Enlarged view of details in the loss-rate measurements in Fig.\ 2(b) of the main text. The blue solid line is the same fit as shown there. (a) $X \leftrightarrow b$ transition. (b) Secondary loss peak. The grey line is a Lorentzian fit with the previously determined lineshape of the main peak added as a background term. This fit which was used to determine the center frequency and linewidth of this peak.}
\label{fig:figS9}
\end{figure}

Data on molecule lifetime in the presence of 866-nm light was obtained by holding molecules in the $| 0 \rangle$ state in a deep 3D lattice. Fig.\ \ref{fig:figS8} shows an example measurement. Fig.\ 2(b) of the main text gives an overview of the data taken in these measurements, while Fig.\ \ref{fig:figS9} shows specific parts in more detail. The molecule association was done at lattice depths of 150 $E_R$ in the vertical direction and 15 $E_R$ to 20 $E_R$ in both horizontal directions. After association, the lattice was ramped to 40 (120) $E_R$ in the vertical (horizontal) direction over \SI{100}{\milli \second} for the data points at detunings $\Delta \geq 2 \pi \times \SI{1}{\giga \hertz}$. This is the lattice configuration in which we observed the longest $1/e$ lifetime of ground-state molecules of \SI{1.4}{\second}. For data points at detunings $\Delta < 2 \pi \times \SI{1}{\giga \hertz}$, no additional lattice ramp was performed, resulting in a molecule lifetime of \SI{0.17}{\second}. The 866-nm beam was then ramped on over a time of 50\,$\mu$s to an intensity of $\SI{1150}{W/cm^2}$ and the molecules were held for various durations before imaging. While inelastic collisions between molecules in the ground band of the lattice are strongly suppressed in deep lattices, there is still heating caused by intensity and frequency noise of the lattice light. This can excite molecules into higher bands where they can move through the lattice more freely, resulting in inelastic close-range collisions between molecules to occur. For each data point in Fig.\ 2(b) of the main text and in Fig.\ \ref{fig:figS9}, the loss rate measured in the respective lattice configuration in absence of 866-nm light was subtracted. 

Figure\ \ref{fig:figS9}(b) shows the frequency range around $\Delta = 2 \pi \times \SI{1.8}{\giga\hertz}$, where we observed an additional loss peak. The properties of this secondary loss feature were determined by fitting a Lorentzian function, to which the known shape of the main feature was added as a background term. This yielded a center frequency at $\Delta = 2 \pi \times \SI{1.78(5)}{\giga\hertz}$ and an effective average linewidth of $ \sqrt{\Gamma\Gamma_e} = 2 \pi \times \SI{185(20)}{\hertz}$. The feature is possibly a transition to one of the rotational states of the $|b^3\Pi_{0^-}, v=0\rangle$ manifold. The splitting between the $|b^3\Pi_{0^+}\rangle$ and $|b^3\Pi_{0^-}\rangle$ potentials can be explained by the coupling of the $\Omega=0^+$ state to the nearby $|A^1\Sigma^+_{0^+}\rangle$ state due to the second order spin-orbit interaction, which affects the $\Omega=0^-$ potential minimally. The transition from $|X^1\Sigma^+_{0^+}\rangle$ to $|b^3\Pi_{0^-}\rangle$ is only allowed because of the mixing with $|b^3\Pi_{0^+}\rangle$ rovibrational states by Coriolis forces or hyperfine interaction depending on the rotational state. Such mixing is usually weak and the transition is difficult
to observe unless the state being mixed with is close by.
\newpage

\end{document}